\documentclass[11pt]{article}

\usepackage[percent]{overpic}
\usepackage{graphicx}
\usepackage{pgfplots}
\usepackage{placeins}
\usepackage{amsmath}
\usepackage{amsthm}
\usepackage{color}
\usepackage{amssymb}
\usepackage{mathtools}
\usepackage{subfigure}
\usepackage{multirow}
\usepackage{listings}
\usepackage{enumitem}
\usepackage{rotating,tabularx}
\usepackage{graphics}
\usepackage{epstopdf}
\usepackage{longtable}
\usepackage[bookmarks=false]{hyperref}
\usepackage{epigraph}
\usepackage{xspace}
\usepackage{amsfonts}
\usepackage{eurosym}
\usepackage{ulem}
\usepackage{url}
\usepackage{footmisc}
\usepackage{comment}
\usepackage{setspace}
\usepackage{geometry}
\usepackage{caption}
\usepackage{pdflscape}
\usepackage{array}
\usepackage[round]{natbib}
\usepackage{booktabs}
\usepackage{dcolumn}
\usepackage{mathrsfs}
\usepackage{booktabs,caption}
\usepackage[flushleft]{threeparttable}
\usepackage[capposition=top]{floatrow}

\usepackage{enumitem}
\usepackage{tikz}
\usetikzlibrary{decorations.pathreplacing}

\usepackage{xcolor}
\hypersetup{
colorlinks,
linkcolor={blue!50!black},
citecolor={blue!50!black},
urlcolor={blue!50!black}}

\newcommand{\bc}{\begin{center}}
\newcommand{\ec}{\end{center}}

\newcolumntype{d}[1]{D{.}{.}{#1}} % "decimal" column type
\newcolumntype{L}[1]{>{\raggedright\let\newline\\arraybackslash\hspace{0pt}}m{#1}}
\newcolumntype{C}[1]{>{\centering\let\newline\\arraybackslash\hspace{0pt}}m{#1}}
\newcolumntype{R}[1]{>{\raggedleft\let\newline\\arraybackslash\hspace{0pt}}m{#1}}

\newcommand{\Sref}[1]{Section~\ref{sec:#1}}

\newcommand{\Cref}[1]{Corollary~\ref{coro:#1}}
\newcommand{\cref}[1]{Cor.~\ref{coro:#1}}

\newcommand{\be}{\begin{equation}}
\newcommand{\ee}{\end{equation}}
\newcommand{\bea}{\begin{eqnarray}}
\newcommand{\eea}{\end{eqnarray}}
\newcommand{\bi}{\begin{itemize}}
\newcommand{\ei}{\end{itemize}}

\numberwithin{equation}{section}

\begin{document}
%\onehalfspacing
\begin{titlepage}
\title{The short-run impact of COVID-19 on the activity in the insurance industry in the Republic of North Macedonia}
\author{Viktor Stojkoski\footnote{Faculty of Economics, University Ss. Cyril and Methodius in Skopje and Macedonian Academy of Sciences and Arts,~\url{vstojkoski@eccf.ukim.edu.mk}}  \and Petar Jolakoski\footnote{Faculty of Economics, University Ss. Cyril and Methodius in Skopje,~\url{jolakoskip@eccf.ukim.edu.mk}}  \and Igor Ivanovski\footnote{Faculty of Economics, University Ss. Cyril and Methodius in Skopje,~\url{igor.ivanovski@eccf.ukim.edu.mk}} }
%\date{First version: August 26, 2018\,\,\,\,\,\,\,\,\,\,\,\,\,\,\,\,\,\,\,\,\,\,\,\,Last revised: \today}
%\date{}
\date{\today}
\maketitle
%\bc
%\red{Preliminary version, please do not circulate}
%\ec
\begin{abstract}
The COVID-19 pandemic had a significant impact on the social and economic actions of every individual. As a consequence, it is expected this impact to transpose into the nature and methods of insuring risky ventures, and thus drastically change the business models of the insurance industry both on short and long run. Despite the abundance of predictions and potential implications, the literature lacks investigations which targets the short-run economic impact of the COVID-19 pandemic on the insurance industry. This paper contributes towards this literature, by performing a first of a kind analysis based on data of the activities in one developing country and insurance market – North Macedonia. By utilizing a seasonal autoregressive model and data on the gross claims paid (GCP) and gross written premiums (GWP) in 11 insurance classes, we evaluate the overall and class-specific impact of the pandemic on the insurance activities in the country. We find that, during the first half of 2020, the activity in GCP and GWP shrank by more than 10\% to what was expected. The total loss in the industry amounted to approximately 8.2 million EUR which was, however, much less than the volume of reserves that the Insurance Supervision Agency made available as funds for the companies to deal with the potential crisis. In addition, the pandemic induced changes in the insurance activity structure - the share of motor vehicles class in the total industry activity fell at the expense of the property classes. Altogether, our results suggest that the insurance industry in North Macedonia was well prepared to tackle the consequences of the pandemic crisis. Nevertheless, the presence of automatic stabilizers had a major influence on weakening the overall impact of the pandemic.

%This paper investigates the impact of the COVID-19 pandemic on the insurance industry in the Republic of North Macedonia during the first half of 2020.  By utilizing seasonal autoregressive models and data for 11 insurance classes, we find that the insurance activity shrank by more than 10\% compared to what was expected.  The total loss in the industry was, however, much less than the amount of funds made available by the Insurance Supervision Agency.  This was because the pandemic induced changes in the activity structure - the share of Motor vehicles class fell at the expense of the property classes.
\end{abstract}
\setcounter{page}{0}
\thispagestyle{empty}
%\nopagebreak
\end{titlepage}
\pagebreak \newpage
%\nopagebreak
\section{Introduction}\label{sec:introduction}

The COVID-19 outbreak led to an unprecedented abrupt economic shock to many developing countries, among which was the Republic of North Macedonia.

In order to reduce the impact of the disease spread, the government of North Macedonia implemented social distancing restrictions such as closure of schools, airports, borders, restaurants and shopping malls. In the most severe cases there were even lockdowns – the citizens of certain municipalities were prohibited from leaving their homes. This subsequently led to a major economic downturn: stock markets plummeted, inter- national trade slowed down, businesses went bankrupt and people were left unemployed.

The resulting pandemic and the government actions taken in response, notably altered the social and economic activities undertaken by the population \citep{stojkoski2020socio}. There is no doubt this will lead to a drastic change in the way economies manage and distribute their risks. As such, it is expected to have a significant impact on the insurance industry, whose goal is to assure people’s activities by developing and supplying products and services for absorption and transferring of risks. The impact of the crisis should translate in the structure of insurers’ claims, assets and business flows.

Even though, many of the pandemic’s consequences will take time to materialize,  and the ultimate effects will depend on the severity of the economic crisis, with preliminary studies suggesting that the insurance industry has remained strong, with insurers generally maintaining their solvency position. As elaborated in a report by \cite{InsuranceEurope}:

``\textit{Insurers across Europe have maintained business continuity, including their ability to continue to serve customers effectively. 
Since the outset of the pandemic, many insurers have taken a very broad range of actions to help their clients. On a case-by-case basis, they have offered, where possible, temporary fee deferrals, fee waivers or even partial refunds of premium payments. Moreover, insurers have launched a broad range of voluntary goodwill actions to support citizens and businesses...}''

The general global losses in the industry are expected to amount to an estimated $84$ billion EUR. Nevertheless, the magnitude of the impact will not be the same in every economy. This is a result of the large heterogeneity in the impact of the pandemic in the economies, as well as the differences in the measures introduced by the governments. In addition, it is also because of to the national features of insurance, that is, the extent to which insurers cover certain losses is dependent on national regulatory and supervisory regimes, besides being dependent on individual company circumstances. 

\subsection{Our contribution}

Definitely, the impact of the pandemic should be studied from a country-specific perspective. To the best of our knowledge, such analyses are yet in their infancy, with so far being made only several contributions, among which \citep{richter2020covid}. In this aspect, here we aim to introduce a coherent statistical procedure of evaluating the short-run impact of the COVID-19 crisis on the activity in the insurance industry, and specifically in the Republic of North Macedonia. North Macedonia is a small Southeastern European developing economy. Such economies are characterized with developing insurance industries and unstable and yet upward trending dynamics of the gross written premiums (GWP) and gross claims paid (GCP) due to the previous lower development, increased liberalization and competition of the market and the constant restructuring of the activities within the insurance classes. GWP are the total revenue from a contract expected to be received by an insurer before deductions for reinsurance or ceding commissions, whereas GCP are all demands made by the insured, for payment of the benefits provided by the insurance contract or for coverage of an incurred loss. Hence, by construction, these two observables are enough to quantify the activity on both the demand and supply side of the insurance industry, i.e., the former represents the demand side, and the latter, implicitly describes the supply side of the industry.

Our paper answers whether, indeed, GWP and GCP decreased (or increased) more than what was expected and to what extent. The paper also clarifies the magnitude of the effect – what were the total losses in the industry? It also addresses the class-specific impact. Precisely, it answers how the gross insurance premiums and gross claims payments in certain classes decreased more than others, and how they affected the overall structure of the insurance industry activities.

To answer these questions, one would have to compare the expected number and volume of claims and premiums to the one realized in the first and second quarter of 2020. While the realized claims and premiums are easily obtainable from the reports of the national Insurance Supervision Agency (ISA), calculating the expected amount might represent an exhausting task. In particular, the expected amount of GWP and GCP is premised on statistical predictions that are made via distinct mathematical models. The choice of model ultimately determines the computational cost and the predictive power -- models that are more computationally expensive also have greater predictive power and vice versa. 

To obtain reliable and inexpensive predictions for the expected value one usually has to make a trade-off and chooses the model which has the best cost-benefit characteristics. As an example, here we consider the Seasonal Auto-regressive Integrated Moving Average model (SARIMA). The advantage of modeling through SARIMA is that it specifies that the output variable depends linearly on its own previous values and on a stochastic term, thus using the best fit recurrence relation to predict future values \citep{makridakis1997arma}. In addition, it is able to capture the possible seasonal pattern which has been observed in the dynamics of GCP and GWP \citep{ulyah2019comparing}.

By utilizing quarterly data, spanning from the second quarter of 2012 up until the last quarter of 2019, we fit a separate SARIMA model for the realized gross insurance premiums and claims for 11 insurance classes. We use the results to construct out of sample forecast for the first and second quarter of 2020 and take the difference of our predictions with the realized values as our measure for the impact of COVID-19.

By performing an analysis of our predictions with respect to the realized values, we report that the activity in GCP and GWP decreased by more than 10 percentages during the first half of 2020, compared to what was expected. These values are much larger when compared with the activity in the same quarter in the previous year, when the changes were between $0$ and $4.5$ percentages. In nominal terms, our predictions suggest the nominal effect of the crisis is around 8.2 million EUR. In comparison, the total non-life and life GWP activity in the first half of 2020 amounted to 172 million EUR. Therefore, on the first sight it might appear that the pandemic had an enormous impact on the activity in the insurance industry. Nevertheless, in contrast to the actions taken by the ISA in order to reduce the impact of the crisis, which were in a value of around 30 million EUR, it seems that the impact was much smaller than what was initially estimated\footnote{The description of the measures introduced by ISA can be found at  \url{http://aso.mk/en/new-set-of-measures-for-insurance-undertakings-to-facilitate-work-in-emergency-conditions/}}.

Besides the overall impact, the COVID-19 pandemic induced a drastic change in the structure of GCP. In particular, we provide evidence that during the first two quarters of 2020 the share of the Property, other and Property, fire and nat. forces classes increased, while at the same time the share of Motor vehicles (Casco) and MTPL classes decreased. We argue that the changes in the structure were a consequence of an automatic stabilizer effect in the risky social and economic undertakings of the population  due to the introduced social distancing measures, even though the MTPL is mandatory and administratively organized. These conclusions, which can not be inferred from a simple descriptive analysis based solely on comparison with past data, may be more consistent with the ``perceived'' economic situation in some countries than otherwise suggested.

The rest of the paper is organized as follows. In \Sref{literature-review} we conduct a thorough review of the literature describing the factors which determine the insurance claims and premiums and discuss their relation with the dynamics of the COVID-19 pandemic. Next, in \Sref{methodology} we describe in detail the SARIMA model and the data used to perform the analysis. \Sref{results} is constituted of three parts. In the first part we provide a descriptive analysis for the dynamics of the gross insurance claims and premiums in the Republic of North Macedonia during the pandemic. We continue by presenting our ARIMA results and quantifying the impact of the COVID-19 crisis. In the last part of this section, we present the implications created by our results. Finally, in \Sref{conclusion} we summarize our findings and offer directions for future research.

\section{Literature Review}\label{sec:literature-review}

The primary contribution of this paper is empirical. It provides new information on the dynamics of the two main insurance activities in North Macedonia following the COVID-19 outbreak. It also enables identifying the differential impact on insurance classes. As such, it joins the growing literature on financial markets impact of an epidemic, focusing mainly on the risky undertakings and the insurance reactions of firms to an epidemic, see for example \citep{acharya2020stress,mcaleer2020prevention,richter2020covid}. The general consensus of this literature is that epidemics can obstruct social and economic welfare by changing expectations about how the economy will function and by deterring investment and tourism. In many situations, the immediate costs of an epidemic are apparent, while the long-term costs are unclear. In any case, it is apparent that epidemics are extremely costly. As noted in \cite{bloom2004epidemics}, preventing epidemics requires overcoming a range of obstacles, as does responding to an epidemic once it begins.

Our contribution also joins the growing literature on how the COVID-19 outbreak specifically contributed to the changes in the activity of the insurance industry \cite{mansour2020outpatient,babuna2020impact,acs2020employment,richter2020covid}. Various approaches have been implemented to investigate the impact of COVID-19 on the insurance industry. For instance, \cite{mansour2020outpatient} used the patient perspective to propose improvements in the coverage of the health insurance class. \cite{babuna2020impact} conducted interviews with representatives from the insurance industry in Ghana and found out that there is a trend of decreasing profits but increasing claims. \cite{richter2020covid} developed a scenario analysis in which they evaluate and summarize the lessons learned from the pandemic crisis by baselining actual developments against a reasonable, pre-COVID-19 scenario. Their results support the hypothesis that financial market developments dominate claims losses due to the demographics of pandemics and other factors. Nevertheless, this literature is still in its infancy and, we believe that, our paper can be a stepping stone towards a more data-driven understanding of the COVID-19 impact on the industry. 

From a methodological perspective the paper contributes to the time-series properties of the insurance activities dynamics, \citet{clinebell1994time,thomann2013impact,arena2008does,ulyah2019comparing,kumar2020forecasting,cummins1985forecasting,mohammadi2013dynamics,andrews2013building}. This literature suggests that GWP and GCP are usually very complex and exhibits seasonal patterns \citet{pesantez2019predicting,choi2009predicting}. Moreover, usually such research on GWP and GCP relies on large scale class specific data, which cannot be integrated together \citet{hong2017flexible,yang2018insurance,guelman2012gradient,diao2019research}. This paper presents a way to use a reliable existing data source that is published on a quarterly basis, the ISA activity reports, to study short-run insurance dynamics. This would allow to easily expand the analysis done here in the future. We also hope that this work will stimulate similar work in other countries with similar data sources.

%Diao, L. and Wang, N., 2019. Research on Premium Income Prediction Based on LSTM Neural Network. Advances in Social Sciences Research Journal, 6(11), pp.256-260.

%Ying, J.J.C., Huang, P.Y., Chang, C.K. and Yang, D.L., 2017, December. A preliminary study on deep learning for predicting social insurance payment behavior. In 2017 IEEE International Conference on Big Data (Big Data) (pp. 1866-1875). IEEE.

%Śmietanka, M., Koshiyama, A. and Treleaven, P., 2020. Algorithms in Future Insurance Markets. Available at SSRN 3641518.

\section{Materials and methods}\label{sec:methodology}

\subsection{Data}

To empirically evaluate the impact of the COVID-19 pandemic on the insurance industry in the Republic of North Macedonia, we use quarterly data for the gross insurance premiums and claims per insurance class.  We focus on the period from the second quarter of 2012, up until the last quarter of 2019. This is the optimal period which allows us to utilize the largest amount of data: including longer periods will result in a smaller longitudinal dataset. During this period, there are a total of 11 insurance classes for which there is no missing data and are included in our analysis. Ten of them belong to the non-life insurance type: Accidents; Health; Motor vehicles (casco); Cargo, Property; fire and nat. forces; Property, other; MTPL (total); General liability; Financial losses and Tourists assistance; whereas one of them belongs to the life insurance type - the Life assurance. The life assurance class incorporates insurances with respect to life, death insurance, mixed life insurance, rent insurance and unit-linked life insurance. From the analysis we exclude the classes for which there are gaps in the reported data and for which the activity is negligible (i.e. most of the time there were either no claims paid or written premiums). Overall, these are a total of 11 classes of insurance. In total the excluded classes account for less than $1\%$ in the activity of GCP and GWP in 2020, and therefore we believe that their exclusion will not significantly impact the interpretation of our findings. The data for the gross insurance premiums and claims were collected from ISA's preliminary reports. Altogether we end
up with 33 observations for each class: 31 of which span the estimation period, while the remaining two quarters represent the basis for quantifying the current impact of the crisis. All data was collected on 28th September 2020 from \url{https://aso.mk/en/preliminary-data/}. The cleared and processed data used in our analysis is available at \url{github.com/pero-jolak/insurance-activity-mkd}.

Table~\ref{tab:summary-stats} gives the summary statistics for each of the insurance classes. During the studied period the MTPL class was the one with most gross claims paid followed by the Motor vehicles (casco) class and the Accident class, whereas the Financial losses class has the lowest number of gross claims paid. Average gross written premiums is also highest for the MTPL (total) class, followed by the Property, other class. Cargo and Financial losses are the classes with the smallest average gross written premiums.

\begin{table}[]
 \begin{threeparttable}
\caption{Summary Statistics \label{tab:summary-stats}}
\begin{tabular}{|l|r|r||r|r|}
\hline
                              & \multicolumn{2}{c||}{GCP} & \multicolumn{2}{c|}{GWP} \\\hline
\textbf{Insurance class}& \textbf{Average}                & \textbf{Std. Dev.}               & \textbf{Average}                &  \textbf{ Std. Dev.}           \\\hline
MTPL (total)                  & 1,025,222           & 477,989 & 2,323,629                & 1,180,925 \\
Financial losses              & 733               & 1,298   & 26,636                  & 18,935   \\
Property, fire and nat.forces & 91,288             & 70,648  & 368,228                 & 151,748  \\
Property, other               & 208,178            & 132,686 & 687,706                 & 266,641  \\
Cargo                         & 2,184              & 1,715   & 47,486                  & 20,767   \\
Motor vehicles (casco)        & 282,362            & 136,530 & 470,010                 & 211,305  \\
Accident                      & 232,642            & 101,869 & 414,355                 & 157,021  \\
General liability             & 25,989             & 25,353  & 129,222                 & 49,824   \\
Tourists assistance           & 27,637             & 15,680  & 110,559                 & 61,139   \\
Health                        & 10,962             & 18,322  & 49,513                  & 58,586   \\
Life assurance                & 120,712            & 80,969  & 650,947                 & 393,785    \\\hline  
\end{tabular}
\begin{tablenotes}
      \small
      \item Note: The data is in 000 MKD. 61.5 MKD = 1 EUR.
      \item Source: Own calculations using data from ISA.
    \end{tablenotes}
\end{threeparttable}
\end{table}

\begin{figure}[!htb]
%\centering
\caption{Dynamics of GCP.\label{fig:bis-dynamics}}
\includegraphics[width=1.0\textwidth]{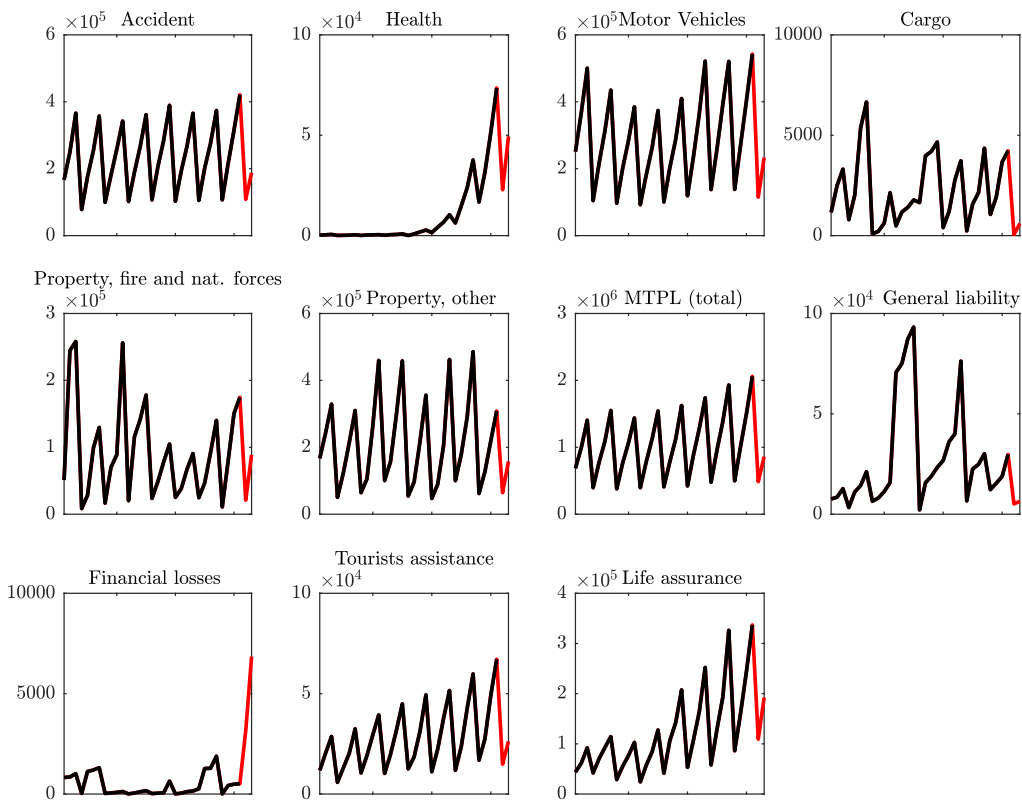}
\floatfoot{Note: On the x-axis is time, measured in quarters, whereas on the y-axis is the value of the gross claims paid in 000 MKD (61.5 MKD = 1 EUR). The red lines correspond to values in 2020, whereas the black lines are values before 2020.}
\end{figure}

\begin{figure}[!htb]
%\centering
\caption{Dynamics of GWP.\label{fig:bpp-dynamics}}
\includegraphics[width=1.0\textwidth]{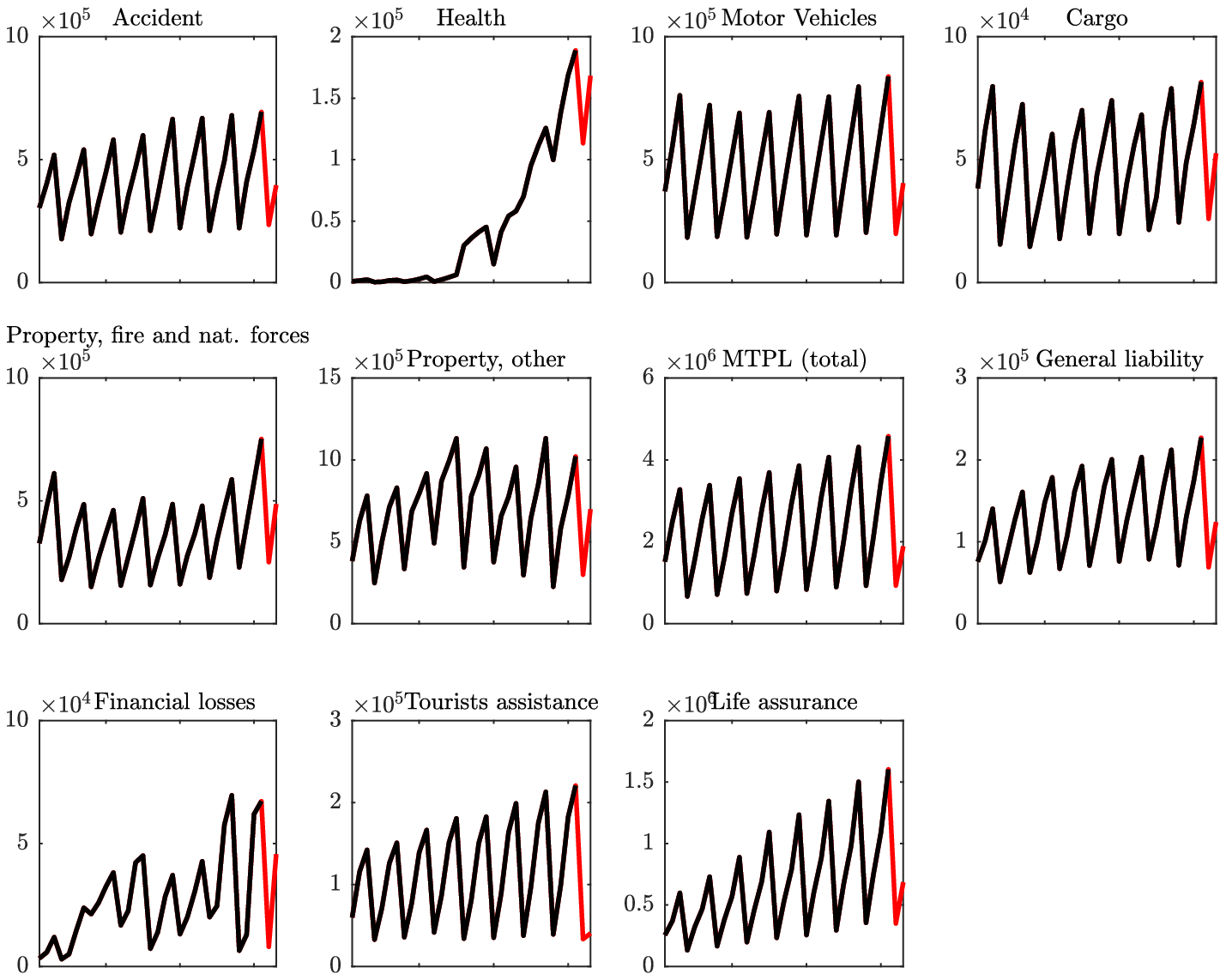}
\floatfoot{Note: On the x-axis is time, measured in quarters, whereas on the y-axis is the value of the gross written premiums in 000 MKD (61.5 MKD = 1 EUR). The red lines correspond to values in 2020, whereas the black lines are values before 2020.}
\end{figure}

Figure~\ref{fig:bis-dynamics} displays the dynamics of the values for the gross claims paid for each class. Every class, except Health and financial losses exhibits a seasonal pattern, thus suggesting the appropriateness of using a seasonal model. The Health and financial losses classes, on the other hand have an upward trend pattern with seasonal adjustments. Hence, under a suitable stationary transformation a seasonal econometric model can be also implemented to them. Similarly, Figure~\ref{fig:bpp-dynamics} displays the dynamics of the values for the gross written premiums for each class. The same dynamical pattern for each class as in the gross claims values appear again, hence a seasonal model can be also implemented for this insurance activity. In the figures, the black lines display the dynamics before 2020, whereas the red lines emphasize the volumes in 2020. For each class and type of activity, we observe that there is a significant drop in the first quarter of 2020, followed by either a stationary or a slight upward trend. Altogether, this indicates that the COVID-19 pandemic had a negative impact on the activity in the insurance industry in North Macedonia.

\subsection{Econometric model}

The simplest, yet effective model for studying seasonal changes is the seasonal autoregressive integrated moving average model (SARIMA). Under this model, a stationary transformation of the data under study, let's name it $x_{i,t}$, where $i$ is a nomenclature for the insurance classes and $t$ is the period of observation, is dependent on its past values, there is a seasonal (periodic) effect and it is determined by the amount of previous shocks in the class\footnote{In the case of GCPs and GWPs, the suitable stationary transformation is given by the logarithm of the reported value}.

Formally, the mathematical specification of an econometric model described solely by an autoregressive and moving average terms is 
\begin{align}
    x_{i,t} = \mu + \phi_1 x_{i,t-1} + \theta_1 \epsilon_{i,t-1} + ... + \phi_p y_{i,t-p} + \theta_q \epsilon_{i,t-q} + \epsilon_{i,t},
\end{align}
where the $\phi_{\hat{i}}$ are the parameters of the marginal effect of the $t-\hat{i}$-th period on the current value of the dependent variable, and  $\theta_{\hat{i}}$ are the respective marginal effects of the previous errors (shocks) $\epsilon_{i,t-\hat{i}}$.
The AR term ($p$) of these models indicates that the evolving variable of interest is modeled as a function of its own prior values. In other words, we have a linear regression setting where the model predicts the next data point by looking at previous data points. The MA (for “moving average”) term ($q$) of the model represents the regression error as a linear combination of the lagged error terms. Finally, $I$ or “integrated” refers to the order of differencing that renders the initial time-series data stationary.

In order to capture the multiplicative effect of seasonal cycles in the data we introduce the SARIMA model. This model is often represented as SARIMA $(p,d,q) \times (P,D,Q)_s$ where the lowercase letters refer to the specification of simple ARIMA model, while the uppercase letters refer to the specification of the seasonal component. Finally, the subscript $s$ indicates the periodicity of the seasons (in our case 4, for quarterly data). Generically, the data process is written as:
\begin{align}
    \phi_p(L)\tilde\phi _P(L^s)x_{i,t} = A(t) + \theta_q(L)\tilde\theta_Q(L^s)\epsilon_t,
\end{align}

where $\phi_p(L)$ and $\tilde\phi_p(L^s)$ are respectively, the non-seasonal and seasonal autoregressive lag polynomials, and $A(t)$ is the intercept. Moreover, the $\Theta_q(L)$ and $\tilde\Theta_Q(L^s)$ are respectively, the non-seasonal and seasonal moving average lag polynomial.

To present the model construction of SARIMA, let us assume that we have selected a SARIMA $(2,1,0) \times (1,1,0)_{4}$ model. After a careful manipulation of the non-seasonal and seasonal lag polynomials we arrive at the following model representation:

\begin{align}
    x_{i,t} = \mu + \phi_1x_{i,t-1}+\phi_2x_{i,t-2}+\tilde\phi_1x_{i,t-4}-\phi_1\tilde\phi_1x_{i,t-5}-\phi_2\tilde\phi_1x_{i,t-6}+\epsilon_t.
\end{align}

Note that the stationary transformation of the variable $x_{i,t}$ represents taking first difference ($d=1$) and then taking the fourth difference ($D=4$). The final model effectively represents an ARIMA model with autoregressive multiplicative effects by taking combinations of the underlying seasonal and non-seasonal parameters.

\subsection{Model selection and diagnostics}

The optimal fit of a particular SARIMA model can be done easily by performing a numerical optimization. The choice of the seasonal, autoregressive and moving average parameters, however, is a far more important task. This is because, in general, it is them which determine the performance of the model.

A standard way for choosing models is by using an information criterion estimator. A such estimator evaluates the relationship between the goodness of fit of the model and its complexity in terms of the number of explanatory variables. Here, we opt to use the Akaike information criterion (AIC), AIC is founded in information theory, and its inference is done by comparing a given a set of candidate models for the data. Always, the preferred model is the one with the minimum AIC value. The inference is done by rewarding goodness of fit, but also including a penalty that is an increasing function of the number of estimated parameters. The penalty discourages overfitting, which is desired because increasing the number of parameters in the model almost always improves the goodness of the fit.

The implementation of AIC allows us to choose a total of 22 distinct SARIMA models, each being able to adequately predict the future GCP and GWP values of a particular insurance class. 

In order to produce consistent predictions, besides providing an adequate fit, the model should also satisfy two assumptions: i) homoscedasticity and ii) absence of autocorrelation between the residuals. To investigate whether our models satisfy these assumptions we conduct two statistical tests. For the first test, we estimate the ARCH LM statistic of each model. The ARCH LM test is the standard approach for detecting whether the model satisfies autoregressive conditional homoscedasticity \citep{engle1982autoregressive}. With the second test, we infer the Ljung-Box statistic \citep{ljung1978measure}. This statistic shows whether there is any autocorrelation between the residuals of every model. Under the null hypothesis, both tests assume that the assumptions are satisfied, and thus the models can be used for predictive purposes.

%ARCH LM test is the standard approach for detecting autoregressive conditional heteroscedasticity.

\section{Results}\label{sec:results}

\subsection{Descriptive analysis}

We begin the analysis by performing a simple comparative analysis for the dynamics of GCP and GWP in the two quarters of 2020. For this purpose, we conduct two distinct evaluations. In particular, first we investigate the differences between the realized values in the first and second quarter of 2020 with the with the ones observed in the previous quarter (last quarter of 2019 and first quarter of 2020, respectively). The quantity which we formally estimate is the single period growth rate of the value of GCP and GWP. Formally, this is given as
\begin{align}
    r_{i,t-1} &= 100 \times \frac{y_{i,t} - y_{i,t-1}}{y_{i,t-1}},
\end{align}
where $y_i(t)$ is the value of either GCP or GWP of class $i$ in time $t$, measured in thousands of MKD (61.5 MKD = 1 EUR). This comparison allows us to examine the trend patterns in the time-series and whether they drastically changed during the pandemic. 

Second, we examine the seasonal patterns during 2020 by comparing the realized values in the two quarters of 2020 with the corresponding quarters of 2019, i.e.,
\begin{align}
    r_{i,t-4} &= 100 \times \frac{y_{i,t} - y_{i,t-4}}{y_i(t-4)}.
\end{align}

Table~\ref{tab:descriptive-analysis-1-quarter} gives the results for the first quarter of 2020. They reveal that the, for the descriptive dynamics of GCP, in the first quarter there was a gain of $0.11$ percentages compared to the same value in the same quarter of the previous year (column base p.y.). Out of the classes, the largest increase was in the Financial losses which exhibited a growth of more than $50000$ percentages, followed by the Property, fire and nat. forces class ($92.74\%$), whereas the GCP in the Cargo class decreased the most during this period ($-94.69\%$). Interestingly, when compared to the previous quarter (column p.q.), only the Financial losses category exhibited, growth whereas every other class had a significant decline in the GCP. The total decline in the activity was $-76.35$ percentages. We believe that this is majorly a consequence of the seasonal patterns present in the GCP dynamics.

Identical conclusions hold for the changes in GWP during the first quarter of 2020. Concretely, the total change in GWP compared to the same values in the first quarter of 2019 is $4.42\%$. The Property, other class showed the largest growth during this period ($32.51\%$), followed by the Financial losses class ($23.73\%$), and the Tourists assistance class was the one with the smallest growth ($-14.76\%$). Every class showed decline in GWP when compared to the previous quarter.

\begin{table}[]
 \begin{threeparttable}
\caption{Percentage changes in the activity in the insurance industry in the 1st quarter of 2020 \label{tab:descriptive-analysis-1-quarter}}
\begin{tabular}{|l|r|r|r|r|}
\hline
& \multicolumn{2}{c||}{GCP} & \multicolumn{2}{c|}{GWP} \\\hline
\textbf{Insurance class}                               & base p.y.        & base p.q.        & base p.y.        & base p.q.      \\\hline
MTPL (total)                  & -2.47             & -76.38          & 0.16                   & -79.67          \\
Financial losses              & 52183.33          & 509.13          & 23.73                  & -88.16          \\
Property, fire and nat.forces & 92.74             & -88.05          & 9.16                   & -66.68          \\
Property, other               & 3.91              & -79.18          & 32.51                  & -70.68          \\
Cargo                         & -94.69            & -98.67          & 5.79                   & -68.22          \\
Motor vehicles (casco)        & -16.69            & -78.8           & -2.51                  & -76.33          \\
Accident                      & 1.51              & -74.26          & 5.97                   & -66.16          \\
General liability             & -58.04            & -82.77          & -3.55                  & -69.67          \\
Tourists assistance           & -11.83            & -77.87          & -14.76                 & -84.92          \\
Health                        & 36.66             & -68.93          & 13.48                  & -40.00          \\
Life assurance                & 26.85             & -67.55          & -1.66                  & -78.23          \\\hline
\textbf{TOTAL}                & \textbf{0.11}     & \textbf{-76.35} & \textbf{4.42}          & \textbf{-75.54} \\\hline
\end{tabular}
\begin{tablenotes}
      \small
      \item Source: Own calculations using data from ISA.
    \end{tablenotes}
\end{threeparttable}
\end{table}

Table~\ref{tab:descriptive-analysis-1-quarter} complements the previous table by displaying the changes in the activity in the insurance industry in the second quarter of 2020. During this quarter there are more drastic changes. In total, the GCP declined by $-6.89\%$ compared to its value in the same quarter in 2019. As in the results for the first quarter, the Financial losses were the class with the largest increase, and the Cargo class had the largest decrease. Differently, from the first quarter of 2019, though, in this case every class exhibited growth in GCP when compared to the previous quarter, and the overall growth in the value of the GCPs was $90.36\%$. When examining the descriptive dynamics of GWP we observe the nearly same behavior -- the overall decline in GWPs compared to the same quarter in the previous year was $-2.51$ percentages and there was an increase of almost $100$ percentages compared the first quarter of 2020.  

\begin{table}[]
 \begin{threeparttable}
\caption{Percentage changes in the activity in the insurance industry in the 2nd quarter of 2020} \label{tab:descriptive-analysis-2-quarter}
\begin{tabular}{|l|r|r|r|r|}
\hline
& \multicolumn{2}{c||}{GCP} & \multicolumn{2}{c|}{GWP} \\\hline
\textbf{Insurance class}                               & base p.y.        & base p.q.        & base p.y.        & base p.q.      \\\hline
MTPL (total)                  & -15.16            & 76.76          & -10.91                 & 103.11         \\
Financial losses              & 1503.28           & 118.23         & 254.69                 & 474.03         \\
Property, fire and nat.forces & 11.82             & 325.19         & 20.65                  & 94.47          \\
Property, other               & 24.54             & 145.36         & 20.33                  & 133.64         \\
Cargo                         & -68.68            & 971.43         & 7.67                   & 102.95         \\
Motor vehicles (casco)        & -13.75            & 102.67         & -4.96                  & 103.85         \\
Accident                      & -12.96            & 73.30           & -3.35                  & 68.53          \\
General liability             & -58.51            & 24.95          & -4.73                  & 80.16          \\
Tourists assistance           & -4.26             & 76.44          & -59.2                  & 20.99          \\
Health                        & 57.10              & 116.18         & 21.99                  & 48.48          \\
Life assurance                & 19.69             & 75.95          & -8.78                  & 96.18          \\\hline
\textbf{TOTAL}                & \textbf{-6.89}    & \textbf{90.36} & \textbf{-2.51}         & \textbf{98.74} \\ \hline
\end{tabular}
\begin{tablenotes}
      \small
      \item Source: Own calculations using data from ISA.
    \end{tablenotes}
\end{threeparttable}
\end{table}

While these results offer a simple depiction for the activity in the insurance industry during the initial COVID-19 pandemic crisis, they fail to highlight the differences between the expected and realized activity in the insurance industry due to three reasons. In particular, the health crisis began at the end of the first quarter, thus the total of the crisis in this period can be only partially evaluate by simply studying the relative changes. Second, even though there is an evident seasonal pattern in the dynamics of most classes, a large amount of them also display a general trend of growth in both GCP and GWP, i.e., the observed changes by the descriptive analysis may be simply a result of this trend. Finally, there is no doubt that the COVID-19 pandemic is reshaping and redefining the social activities within an economy. In this aspect, it is obvious that an analysis that does not capture the expected dynamics and is instead strictly rooted in naive predictions will fail to be useful as a coherent tool for developing economic policies.

In spite of all of its shortcomings and due to their simple interpretation, the findings from the analysis in this part can be still used as a baseline, though with a dose of caution, for identifying the implications created by the SARIMA analysis performed in the following part.

\subsection{Measuring the COVID-19 pandemic impact}

As described in \Sref{methodology}, we use the SARIMA model to measure the COVID-19 crisis impact on the activity in the insurance industry. This is done by fitting a separate model to each class and type of activity, based on the data from the 2nd quarter of 2012 up until the last quarter of 2019. Appendix~\ref{sec:appendix-sarima} provides tables with detailed information regarding the choice and performance of the models in fitting our studied data. Each model satisfies the homoscedasticity and no autocorrelation assumptions.

We utilize the models to predict the activity in the 1st and 2nd quarter of 2020 and use the expected value of the prediction $y_i^p(t)$ as our assessment for the expected behavior of the activity in the insurance industry. Subsequently, we use this expected value to estimate the relative percentage error, defined as
\begin{align}
    r_i^p(t) &= 100\times\frac{y_i^p(t)-y_i(t)}{y_i(t)}.
\end{align}
This is our measure for the COVID-19 pandemic impact.

Table~\ref{tab:sarima-relative-errors} displays the results for each class divided per quarter and per type of activity. The total relative error in the GCP activity was $5.29\%$ in the first quarter of 2020 and $11.69\%$. The positive sign implies that, in fact, during the first half of 2020 the GCPs decreased more than expected. The decrease was of similar magnitude in both quarters. Nevertheless, one has to keep in mind that the COVID-19 pandemic began at the end of the first quarter, and hence, only a part of the differences should be attributed to it.

\begin{table}[]
 \begin{threeparttable}
\caption{Relative error in predicted changes in the activity in the insurance industry in 2020. \label{tab:sarima-relative-errors}}
\begin{tabular}{|l|r|r|r|r|}
\hline
& \multicolumn{2}{c|}{GCP} & \multicolumn{2}{c|}{GWP} \\\hline
\textbf{Insurance  Class}                            & \multicolumn{1}{|c|}{1Q}              & \multicolumn{1}{|c|}{2Q}             & \multicolumn{1}{|c|}{1Q}            & \multicolumn{1}{|c|}{2Q}            \\\hline
MTPL (total)                  & 9.09        & 24.90     & 5.56       & 18.46      \\
Financial losses              & -97.74      & -95.42    & 164.36     & -31.70     \\
Property, fire and nat.forces & -29.11      & -7.39     & 16.18      & 4.16       \\
Property, other               & -33.20      & -37.86    & -26.95     & -22.07     \\
Cargo                         & 6136.45     & 358.77    & -3.61      & -6.03      \\
Motor vehicles (casco)        & 24.76       & 20.19     & 6.56       & 8.31       \\
Accident                      & 3.64        & 17.41     & -4.43      & 3.50       \\
General liability             & 318.55      & 202.35    & 14.91      & 14.56      \\
Tourists assistance           & 23.19       & 10.96     & 21.68      & 155.80     \\
Health                        & 22.84       & 20.07     & 103.90     & 66.44      \\
Life assurance                & -22.27      & -19.51    & 12.10      & 18.65      \\ \hline
\textbf{TOTAL}                & \textbf{5.29}   & \textbf{11.69}  & \textbf{8.12}   & \textbf{11.32} \\\hline
\end{tabular}
\begin{tablenotes}
      \small
      \item Source: Own calculations using data from ISA.
    \end{tablenotes}
\end{threeparttable}
\end{table}

In order to investigate the drivers of the differences between the expected and observed GCP activity we inspect the class specific behavior. In this case, in the first quarter we observe large disparities in the behavior: there were 4 classes which experienced activity greater than expected GCP and 7 classes whose GCP activity decreased. Remark that a negative sign in the GCP relative error implies that the expected GCP were smaller than the realized ones. The class whose expected activity decreased mostly than expected was Cargo, whereas the Financial losses was the one with the largest increase in the activity. In the second quarter the same classes show increase in the activity that is bigger than what was expected, and the Financial losses class was again the one with the largest increase. In this quarter, it is also the Cargo class which experienced the largest discrepancies in the GCP activity.

Regarding the GWP activity, the total relative error in the first quarter is $8.12\%$ and in the second quarter it is $11.32\%$. In the case of GWP, the positive value of the relative error indicates that there was less activity than what was expected. The magnitude of the relative error is significant, therefore suggesting that the COVID-19 pandemic had a large impact on the reduction in the overall activity of GWP in the economy of North Macedonia.

When we look at the intra-class errors,  we find that even much larger discrepancies. Specifically, in the first quarter it was the Financial losses class whose activity decreased significantly, whereas in the second quarter it is the Tourists assistance class which decreased mostly. On the other hand, it was the Property, other class and the Financial losses class whose GWP was much more than expected in the first, and second quarter respectively.

In Appendix~\ref{sec:appendix-confidence-intervas} we show the confidence intervals of our predictions. They appear tight, and thus we can infer that our results are robust.

To summarize, these findings allow us to conclude that, overall, the COVID-19 pandemic might have had a serious impact on the activities in the insurance industry in North Macedonia. In absolute terms, there were 0.8 million EUR less paid in GCP during the first quarter of 2020 and 3.4 million EUR less in the second quarter of 2020. In the same fashion, there were a total of 3.3 million EUR less spent in purchases of GWP in the first quarter of 2020, and 9.2 million EUR less in the second quarter of 2020. Altogether, this implies that the loss in the insurance industry can be quantified as a total of 8.2 million EUR. In comparison, ISA produced a set of measures in which it included a release 30 million EUR in liquid assets whose main purpose is to strengthen the support towards the companies in the segment of managing the negative consequences of the COVID-19 crisis. On the one hand, might suggest that the potential impact of the crisis was less than it was actually estimated by ISA. On the other hand, it implies that the insurance industry in North Macedonia was well prepared for the emergence of a such crisis.

\subsection{Implications}
While the COVID-19 crisis had a major impact on the overall activity in the insurance industry, it also induced a significant impact in the class-specific behavior. An exact interpretation of our results requires a detailed investigation of the structure of GCP and GWP in 2020. As a means to provide a such analysis, we compare the expected share of each insurance class in the total expected activity of GCP and GWP with respect to the realized one. 

\begin{figure}[!htb]
%\centering
\caption{Expected and realized share of the classes in the total insurance activity .\label{fig:activity-shares}}
\includegraphics[width=1.0\textwidth]{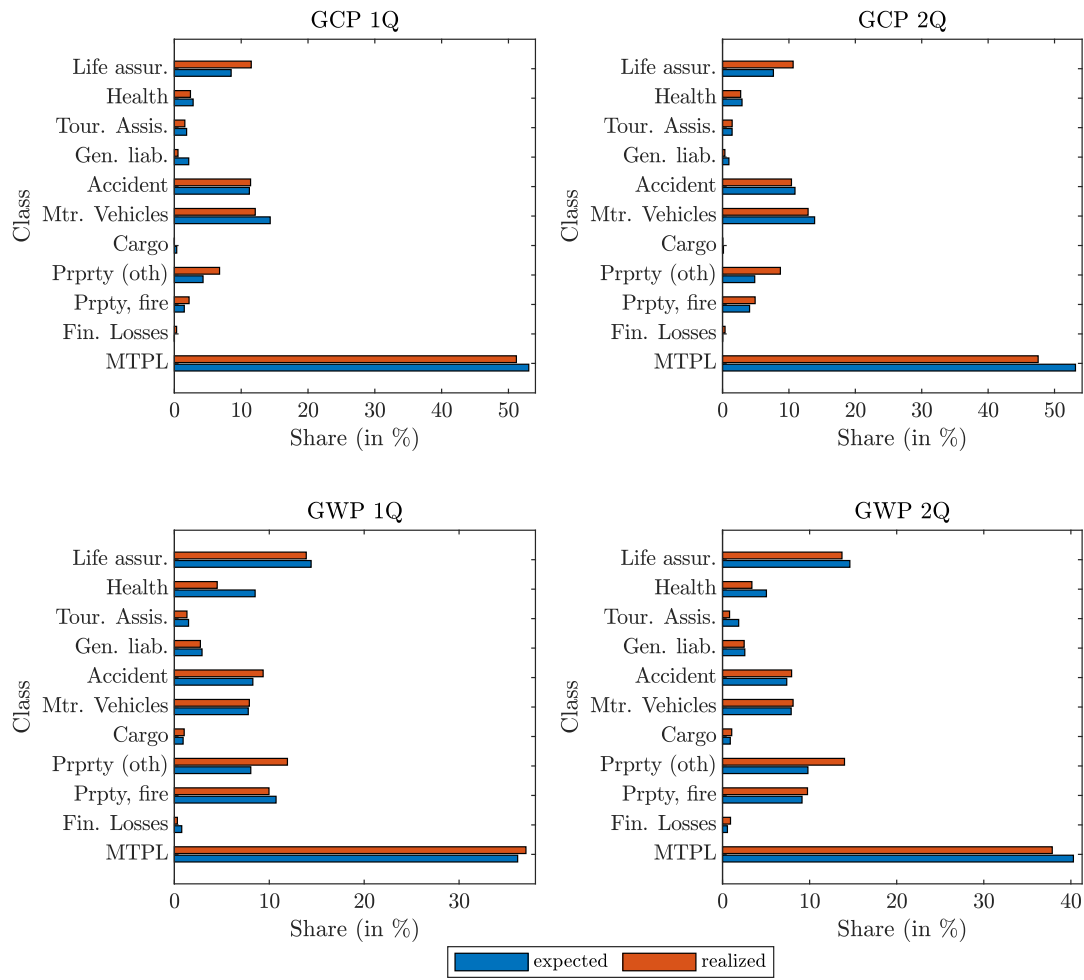}
\end{figure}

The results are presented in Figure~\ref{fig:activity-shares} Evidently, there were several major changes in the structure of the GCP. During the first half of 2020, it was the MTPL and Motor vehicles classes whose share decreased the most. In particular, we discover that in the first quarter of 2020 the share of the these two classes in the total GCP was around $2$ percentage points less. At the same time, it was the Life assurance class whose share increased most, by $9$ percentage points more than expected in the first quarter, by $3$ percentage points. In the second quarter,  we observe an even larger decrease in the MTPL (total) class in the first quarter ($6$ percentage points), and a large increase in the Property, other class ($4$ percentage points). 

In general, it is predicated that the insurance activity in a developing economy, such as the one of North Macedonia, follows an easily predictable pattern \citep{arena2008does}. Hence, it seems that during the 2020 pandemic crisis there was a drastic change in the structure of the GCP. But why was this the case?

The pandemic forced drastic changes in the social and economic activities of the whole population. In particular, there were the government imposed social distancing measures such as movement and travel restrictions, people were instructed to work from home and closure of public places. Moreover, most people were cautious and further decreased the social activities by self-imposing additional distancing measures. In this aspect, it is obvious that vehicle insurance is usually purchased for travels, and this activity was significantly reduced especially during the second quarter of 2020. At the same time, the social distancing measures increased the risky ventures underdone in the home of the insured person. Hence, there has been an increase in GCP share of the Property classes.

%Unfortunately, the COVID-19 pandemic increased the mortality rate in the country, and as a consequence we observe the larger than expected growth in the claims paid. 

Similar changes appear in the structure of GWP. That is, we observe extreme increase in the amount of premiums paid in the Property, other class in the second quarter of 2020, whereas the Health and MTPL classes display a decrease. %There was no other class whose realized share in the total GWP activity increased or decreased more than $1$ percentage point in the both quarters of 2020.

As a result, we can conjecture that the magnitude of the impact of the COVID-19 pandemic was reduced simply because of automatic stabilizer effects that offset the fluctuations in the insurance activity through their normal operation without additional timely authorization by the government or policymakers.
 
\section{Conclusion}\label{sec:conclusion}

We investigated the impact of the COVID-19 pandemic crisis on the activity in the insurance industry in North Macedonia. By including quarterly data on gross claims paid and gross written premiums for 11 classes of insurance, we introduced predictions for the expected behavior of the activity in GCP and GWP in the absence of the pandemic. We measured the impact of the crisis as the difference between our predictions and the realized values in the first and second quarter of 2020.

Similarly to previous works dealing with the dynamics of the insurance activity within an economy, we described the dynamics of the activities of each class as a seasonal autoregressive model. This resulted in a total of 22 SARIMA models, each predicting the gross claims paid of the gross premiums purchased of a particular class.

Our analysis showed that, the GCP activity decreased by $5.29$ percentages in the first quarter and by $11.69$ percentages in the second quarter, compared to what was expected. At the same time, the GWP activity fell by $8.12\%$ in the first quarter and by $11.32\%$ in the second quarter. In comparison to the behavior with the same quarter in the previous year, we observed notable differences. This led us to believe that the COVID-19 pandemic had a significant impact on the insurance activity in North Macedonia. Importantly, we estimated that in nominal value the loss in the 11 studied insurance classes amounted to approximately 8.2 million EUR. This is less than one third of the reserves which ISA made available to the insurance companies in order to increase their liquidity during the crisis. Thus, we argued that the insurance industry was ready to intervene in case of an even larger crisis.

Next, we investigated the structural changes in the activity of the insurance industry during the crisis. We found that the pandemic induced drastic changes in the structure of GCPs and GWPs. In particular, the share of the Property classes in the total activity increased, at the expense of the shares of the vehicle insurance classes. We argued that this was a result of the automatic stabilizers inducing changes due to the introduced social distancing measures. 

In this aspect, it remains an unresolved question why the automatic stabilizers had a such big effect on the impact of the crisis. We believe that the answer to this question lies in the insurance protection gap in the country \citep{richter2020covid}. Formally, as defined in \cite{schanz2018understanding}, the insurance protection gap is the difference between the amount of insurance that is economically beneficial and the amount of coverage actually purchased. It is believed that in developing and emerging countries, such as North Macedonia, this gap is largest because combined insurance premiums still fall significantly short of these countries’ and regions’ share in global GDP. A larger protection gap makes the structure of the industry more susceptible to shocks in case of extreme events. 

In North Macedonia, the exact magnitude of the protection gap is still unknown. A more detailed analysis is certainly needed to uncover the effect of the protection gap on the activity in the insurance industry in North Macedonia, especially in the times of the COVID-19 pandemic and its aftermath. This, however, requires a highly accurate and disaggregated data for the affordability of the insurance policies, their quality and the cultural and social character of the insurer behavior. In regard to this, building an explanatory model for the underlying protection gap would bring novel insights about the relations between the insurance classes in the GCP and GWP and, importantly, about the policies required for fast and stable recovery of the insurance industry from the ongoing crisis. This is a subject of our current work and it aims to offer new recommendations for promoting growth and development of the insurance market.

%\bibliography{insurance}

% Bibstyle aea.bst version 2009.05.20

\appendix

\setcounter{equation}{0}
\setcounter{figure}{0}
\setcounter{table}{0}
\setcounter{theorem}{0}
\makeatletter
\renewcommand{\theequation}{A\arabic{equation}}
\renewcommand{\thetable}{A\arabic{table}}
\renewcommand{\thefigure}{A\arabic{figure}}
\renewcommand{\thetheorem}{A\arabic{theorem}}
\renewcommand{\theproposition}{A\arabic{proposition}}

\newpage

\section{SARIMA model \label{sec:appendix-sarima}}

Table~\ref{tab:gcp_models} and~\ref{tab:gwp_models} show the 22 SARIMA models for each class and the two types of activities that had the best performance under the AIC criterion.

\begin{table}[]
\caption{Selected GCP models and diagnostics. \label{tab:gcp_models}}
\resizebox{\textwidth}{!}{%
\begin{tabular}{|l|l|l|l|l|} \hline
Insurance class &
  Model &
  AIC &
  \begin{tabular}[c]{@{}l@{}}ARCH-LM \\      (p-value)\end{tabular} &
  \begin{tabular}[c]{@{}l@{}}Ljung-Box \\      (p-value)\end{tabular} 
   \\ \hline

MTPL (total)      & SARIMA (1, 0, 0)x(2, 1, 0, 4) & -100.29 & 0.62 & 0.04 \\
Financial losses  & SARIMA (1, 1, 3)x(0, 0, 0, 0) & 130.38  & 0.67 & 0.13 \\
Property, fire and nat.forces &
  SARIMA (3, 1, 3)x(0, 0, 0, 0) &
  54.06 &
  0.24 &
  0.69 \\
Property, other   & SARIMA (1, 0, 0)x(2, 1, 0, 4) & 2.93    & 0.99 & 0.27 \\
Cargo             & SARIMA (1, 1, 4)x(0, 0, 0, 0) & 95.06   & 0.44 & 0.31 \\
Motor vehicles (casco) &
  SARIMA (1, 0, 0)x(0, 1, 0, 4) &
  -99.72 &
  0.34 &
  0.96 \\
Accident          & SARIMA (1, 0, 0)x(0, 1, 0, 4) & -80.25  & 0.85 & 1.00 \\
General liability & SARIMA (1, 0, 0)x(0, 0, 0, 0) & 84.81   & 0.76 & 0.86 \\
Tourists assistance &
  SARIMA (1, 0, 0)x(2, 1, 0, 4) &
  -25.68 &
  0.78 &
  0.38 \\
Health            & SARIMA (3, 1, 3)x(0, 0, 0, 0) & 48.67   & 0.04 & 0.94 \\
Life assurance    & SARIMA (1, 0, 0)x(2, 1, 0, 4) & -19.31  & 0.85 & 0.84 \\ \hline
\end{tabular}%
}
\end{table}
\begin{table}[]
\caption{Selected GWP models and diagnostics. \label{tab:gwp_models}}
\resizebox{\textwidth}{!}{%
\begin{tabular}{|l|l|l|l|l|} \hline
Insurance class &
  Model &
  AIC &
  \begin{tabular}[c]{@{}l@{}}ARCH-LM \\      (p-value)\end{tabular} &
  \begin{tabular}[c]{@{}l@{}}Ljung-Box \\      (p-value)\end{tabular} \\ \hline
MTPL (total)           & SARIMA (0, 1, 2)x(2, 1, 0, 4) & -180.22 & 0.56 & 0.05 \\
Financial losses       & SARIMA (3, 1, 3)x(0, 0, 0, 0) & 49.26   & 0.39 & 0.05 \\
Property, fire and nat.forces &
  SARIMA (1, 0, 0)x(0, 1, 0, 4) &
  -85.39 &
  0.84 &
  0.38 \\
Property, other        & SARIMA (0, 1, 2)x(2, 1, 0, 4) & -20.02  & 0.48 & 0.91 \\
Cargo                  & SARIMA (1, 0, 0)x(0, 1, 0, 4) & -43.35  & 0.33 & 0.52 \\
Motor vehicles (casco) & SARIMA (1, 0, 0)x(0, 1, 0, 4) & -125.19 & 0.54 & 0.89 \\
Accident               & SARIMA (2, 0, 0)x(1, 1, 0, 4) & -105.34 & 0.93 & 0.20 \\
General liability      & SARIMA (0, 1, 2)x(2, 1, 0, 4) & -102.45 & 0.19 & 0.96 \\
Tourists assistance    & SARIMA (0, 1, 2)x(2, 1, 0, 4) & -78.39  & 0.83 & 0.98 \\
Health                 & SARIMA (1, 1, 1)x(0, 0, 0, 0) & 67.20   & 0.43 & 0.70 \\
Life assurance         & SARIMA (0, 1, 2)x(2, 1, 0, 4) & -125.70 & 0.51 & 0.58 \\ \hline             
\end{tabular}%
}
\end{table}

\setcounter{equation}{0}
\setcounter{figure}{0}
\setcounter{table}{0}
\setcounter{theorem}{0}
\makeatletter
\renewcommand{\theequation}{B\arabic{equation}}
\renewcommand{\thetable}{B\arabic{table}}
\renewcommand{\thefigure}{B\arabic{figure}}
\renewcommand{\thetheorem}{B\arabic{theorem}}
\renewcommand{\theproposition}{B\arabic{proposition}}

\section{Confidence intervals of the predicted values \label{sec:appendix-confidence-intervas}}

Tables~\ref{tab:sarima-gcp-1q-estimations},~\ref{tab:sarima-gcp-2q-estimations}~\ref{tab:sarima-gwp-1q-estimations} and ~\ref{tab:sarima-gwp-2q-estimations} show the upper and lower confidence interval bounds for the predicted values of our models. In general, the total upper and lower values of the intervals do not significantly differ from the predicted value, and thus we conclude that our results offer a consistent depiction of the expected activity in the insurance industry in case of no pandemic.

\begin{table}[]
 \begin{threeparttable}
\caption{Prediction and realized values for GCP in the first quarter of 2020. \label{tab:sarima-gcp-1q-estimations}}
\begin{tabular}{|l|r|r|r|r|r|}
\hline
Insurance class & \multicolumn{1}{r|}{95\% CI (lower)} & \multicolumn{1}{r|}{Expected} & \multicolumn{1}{r|}{95\% CI (upper)} & \multicolumn{1}{r|}{Real} & \multicolumn{1}{r|}{Difference} \\ \hline
MTPL (total)                  & 488,683         & 530,810   & 576,569        & 486,558 & -44,252    \\
Financial losses              & 3               & 71        & 1,743          & 3,137   & 3,066      \\
Property, fire and nat.forces & 6,165           & 14,807    & 35,562         & 20,887  & 6,080      \\
Property, other               & 27,730          & 42,928    & 66,454         & 64,263  & 21,335     \\
Cargo                         & 506             & 3,492     & 24,093         & 56      & -3,436     \\
Motor vehicles (casco)        & 131,294         & 143,517   & 156,878        & 115,033 & -28,484    \\
Accident                      & 99,434          & 112,320   & 126,876        & 108,374 & -3,946     \\
General liability             & 3,957           & 21,459    & 116,362        & 5,127   & -16,332    \\
Tourists assistance           & 13,901          & 18,310    & 24,118         & 14,863  & -3,447     \\
Health                        & 12,576          & 28,068    & 62,642         & 22,849  & -5,219     \\
Life assurance                & 62,633          & 84,995    & 115,343        & 109,341 & 24,346     \\ \hline
\textbf{TOTAL} & \textbf{846,882}	& \textbf{1,000,777}	& \textbf{1,306,640}	& \textbf{950,488}	& \textbf{-50,289} \\ \hline

\end{tabular}
\begin{tablenotes}
      \small
      \item Notes: The lower and upper columns show the $95\%$ bounds of the confidence intervals, the expected column is our prediction, the real column is the observed value of the class specific activity and the diff column calculates the differences between the realized and expected value. The values are in in 000 MKD (61.5 MKD = 1 EUR).
    \end{tablenotes}
\end{threeparttable}
\end{table}

\begin{table}[]
\begin{threeparttable}
\caption{Prediction and realized values for GCP in the second quarter of 2020. \label{tab:sarima-gcp-2q-estimations}}
\begin{tabular}{|l|r|r|r|r|r|}
\hline
\multicolumn{1}{|l|}{Insurance class} &
  \multicolumn{1}{r|}{95\% CI (lower)} &
  \multicolumn{1}{r|}{Expected} &
  \multicolumn{1}{r|}{95\% CI (upper)} &
  \multicolumn{1}{r|}{Real} &
  \multicolumn{1}{r|}{Difference} \\ \hline
MTPL (total)                  & 975,619         & 1,074,204 & 1,182,752       & 860,049   & -214,155   \\
Financial losses              & 8               & 313       & 11,592          & 6,846     & 6,533      \\
Property, fire and nat.forces & 33,074          & 82,249    & 204,537         & 88,810    & 6,561      \\
Property, other               & 53,978          & 97,979    & 177,848         & 157,677   & 59,698     \\
Cargo                         & 196             & 2,753     & 38,566          & 600       & -2,153     \\
Motor vehicles (casco)        & 248,111         & 280,212   & 316,467         & 233,135   & -47,077    \\
Accident                      & 193,140         & 220,511   & 251,761         & 187,812   & -32,699    \\
General liability             & 3,294           & 19,369    & 113,889         & 6,406     & -12,963    \\
Tourists assistance           & 21,050          & 29,098    & 40,222          & 26,224    & -2,874     \\
Health                        & 19,258          & 59,309    & 182,651         & 49,394    & -9,915     \\
Life assurance                & 99,852          & 154,856   & 240,161         & 192,388   & 37,532     \\ \hline
\textbf{TOTAL} & \textbf{1,647,580}	& \textbf{2,020,853}	& \textbf{2,760,448}	& \textbf{1,809,341}	& \textbf{-211,512} \\ \hline

\end{tabular}
\begin{tablenotes}
      \small
      \item Notes: The lower and upper columns show the $95\%$ bounds of the confidence intervals, the expected column is our prediction, the real column is the observed value of the class specific activity and the diff column calculates the differences between the realized and expected value. The values are in in 000 MKD (61.5 MKD = 1 EUR).
    \end{tablenotes}
\end{threeparttable}
\end{table}

\begin{table}[]
 \begin{threeparttable}
\caption{Prediction and realized values for GWP in the first quarter of 2020. \label{tab:sarima-gwp-1q-estimations}}
\begin{tabular}{|l|r|r|r|r|r|}
\hline
\multicolumn{1}{|l|}{Insurance class} &
  \multicolumn{1}{r|}{95\% CI (lower)} &
  \multicolumn{1}{r|}{Expected} &
  \multicolumn{1}{r|}{95\% CI (upper)} &
  \multicolumn{1}{r|}{Real} &
  \multicolumn{1}{r|}{Difference} \\ \hline
MTPL (total)                  & 960,985         & 982,417   & 1,004,326       & 930,651   & -51,766    \\
Financial losses              & 9,348           & 21,033    & 47,323          & 7,956     & -13,077    \\
Property, fire and nat.forces & 260,102         & 290,976   & 325,515         & 250,442   & -40,534    \\
Property, other               & 163,220         & 218,617   & 292,816         & 299,261   & 80,644     \\
Cargo                         & 20,034          & 24,988    & 31,167          & 25,924    & 936        \\
Motor vehicles (casco)        & 199,289         & 211,408   & 224,264         & 198,391   & -13,017    \\
Accident                      & 208,045         & 224,523   & 242,307         & 234,936   & 10,413     \\
General liability             & 73,219          & 79,105    & 85,464          & 68,839    & -10,266    \\
Tourists assistance           & 36,144          & 40,508    & 45,399          & 33,291    & -7,217     \\
Health                        & 67,385          & 231,079   & 792,431         & 113,329   & -117,750   \\
Life assurance                & 371,055         & 391,308   & 412,667         & 349,066   & -42,242    \\ \hline
\textbf{TOTAL} & \textbf{2,368,825}	& \textbf{2,715,962}	& \textbf{3,503,679}	& \textbf{2,512,086}	& \textbf{-203,876} \\ \hline
\end{tabular}

\begin{tablenotes}
      \small
      \item Notes: The lower and upper columns show the $95\%$ bounds of the confidence intervals, the expected column is our prediction, the real column is the observed value of tha class specific activity and the diff column calculates the differences between the realized and expected value. The values are in in 000 MKD (61.5 MKD = 1 EUR).
    \end{tablenotes}
\end{threeparttable}
\end{table}

\begin{table}[]
 \begin{threeparttable}
\caption{Prediction and realized values for GWP in the second quarter of 2020. \label{tab:sarima-gwp-2q-estimations}}
\begin{tabular}{|l|r|r|r|r|r|}
\hline
\multicolumn{1}{|l|}{Insurance class} &
  \multicolumn{1}{r|}{95\% CI (lower)} &
  \multicolumn{1}{r|}{Expected} &
  \multicolumn{1}{r|}{95\% CI (upper)} &
  \multicolumn{1}{r|}{Real} &
  \multicolumn{1}{r|}{Difference} \\ \hline
MTPL (total)                  & 2,184,572       & 2,239,191 & 2,295,176       & 1,890,229 & -348,962   \\
Financial losses              & 9,610           & 31,192    & 101,242         & 45,670    & 14,478     \\
Property, fire and nat.forces & 434,212         & 507,319   & 592,736         & 487,044   & -20,275    \\
Property, other               & 395,523         & 544,882   & 750,643         & 699,201   & 154,319    \\
Cargo                         & 38,217          & 49,442    & 63,964          & 52,614    & 3,172      \\
Motor vehicles (casco)        & 406,710         & 438,030   & 471,762         & 404,421   & -33,609    \\
Accident                      & 373,338         & 409,792   & 449,804         & 395,928   & -13,864    \\
General liability             & 131,396         & 142,078   & 153,629         & 124,022   & -18,056    \\
Tourists assistance           & 89,650          & 103,032   & 118,412         & 40,278    & -62,754    \\
Health                        & 68,963          & 280,083   & 1,137,509       & 168,275   & -111,808   \\
Life assurance                & 764,619         & 812,527   & 863,437         & 684,787   & -127,740   \\ \hline
\textbf{TOTAL} & \textbf{4,896,810}	& \textbf{5,557,569}	& \textbf{6,998,314}	& \textbf{4,992,469}	& \textbf{-565,100} \\ \hline

\end{tabular}
\begin{tablenotes}
      \small
      \item Notes: The lower and upper columns show the $95\%$ bounds of the confidence intervals, the expected column is our prediction, the real column is the observed value of tha class specific activity and the diff column calculates the differences between the realized and expected value. The values are in in 000 MKD (61.5 MKD = 1 EUR).
    \end{tablenotes}
\end{threeparttable}

\end{table}

\end{document}